\documentclass[12pt]{aastex}
\usepackage{emulateapj5}
\usepackage{graphicx}
\usepackage{times,psfig}

\def\ltsima{$\; \buildrel < \over \sim \;$}
\def\lsim{\lower.5ex\hbox{\ltsima}}
\def\gtsima{$\; \buildrel > \over \sim \;$}
\def\gsim{\lower.5ex\hbox{\gtsima}}

\newcommand{\be}{\begin{equation}}
\newcommand{\en}{\end{equation}}
\newcommand{\ergs}{\rm \ erg \; s^{-1}}

\def\cmdue {\rm \ cm^{-2}}

\def\ppm{$\pm$}

\begin{document}

\received{~~} \accepted{~~}
\journalid{}{}
\articleid{}{}

\title{The evolution of the high energy tail in the quiescent spectrum of the
soft X--ray transient Aql X-1} 

\author{S.~Campana\altaffilmark{1}, L. Stella\altaffilmark{2}}

\altaffiltext{1}{INAF-Osservatorio Astronomico di Brera, Via Bianchi 46, I--23807
Merate (Lc), Italy}

\altaffiltext{2}{INAF-Osservatorio Astronomico di Roma,
Via Frascati 33, I--00040 Monteporzio Catone (Roma), Italy}


\begin{abstract}
A moderate level of variability has been detected in the quiescent luminosity
of several neutron star soft X--ray transients. Spectral variability was first
revealed by Chandra observations of Aql X-1 in the four months that followed
the 2000 X--ray outburst. By adopting the canonical model for quiescent
spectrum of soft X--ray transients, i.e. an absorbed neutron star atmosphere
model plus a power law tail, Rutledge et al. (2002a) concluded that the
observed spectral variations can be ascribed to temperature variations of the
neutron star atmosphere. These results can hardly be reconciled with the  
neutron star cooling that is expected to take place in between outbursts
(after deep crustal heating in the accretion phase).  
Here we reanalyse the Chandra spectra of Aql X-1, together with a long
BeppoSAX observation in the same period, and propose a different
interpretation of the spectral variability: that this is due to correlated
variations of the power law component and the column density ($>5$, a part of which
might be intrinsic to the source), while the temperature and flux of the
neutron star atmospheric component remained unchanged. 
This lends support
to the idea that the power law component arises from emission at the shock
between a radio pulsar wind and inflowing matter from the companion star.  
\end{abstract}

\section{Introduction}

The large luminosity swing of transient X--ray binaries allows the sampling a
variety of physical conditions that are inaccessible to accreting compact
objects in persistent sources. The very low luminosity that characterises the
quiescent state of neutron star soft X--ray transients, SXRTs, ($L_X \sim
10^{32}-10^{33}\ergs$) opens up the possibility of studying these old, fast
spinning neutron stars in different and yet unexplored regimes such as
accretion onto the neutron star magnetosphere (propeller), resumed millisecond
radio pulsar activity, and/or low-level atmospheric emission from the cooling 
of the neutron star in between the accretion intervals of the outbursts
(e.g. Campana et al. 1998a; Brown et al. 1998; Rutledge et al. 2002b). 

In recent years the quiescent properties of a handful SXRTs have been studied 
in some detail.
The main outcome of these investigations is that the quiescent X--ray spectra
of SXRTs display a soft component plus a hard (power-law) component
contributing a comparable flux in the 0.5--10 keV band (Campana 2001;
Bildsten \& Rutledge 2000; Wijnands 2001).  
The soft component has been frequently modelled with a black body model of
0.1--0.3 keV temperature and few km radius. Especially promising is the idea
that the soft component of SXRTs may be produced from the cooling of the
neutron star heated during the repeated outbursts (van Paradijs et al. 1987;
Stella et al. 1994; Campana et al 1998a). The theory of deep crustal heating
by pycnonuclear reactions compares well with the observations (Brown et
al. 1998; Campana et al. 1998a; Rutledge et al. 1999; Colpi et al. 2001). In
particular, Rutledge et al. (1999) fitted neutron star atmospheric models to
the soft component of quiescent spectra of SXRTs and derived slightly smaller
temperatures (0.1--0.3 keV) and larger radii (10--15 km, consistent with
the neutron star radius) than those inferred from simple black body fits.

Observationally, the hard component is well described by a power law tail.
In the quiescent spectrum of Aql X-1 and Cen X-4 observed by ASCA and BeppoSAX
this component is statistically significant (Asai et al. 1996, 1998; Campana
et al. 1998b, 2000) with photon index in the 1--2 range. The same power law
is needed (even if not statistically significant) in the analysis of Chandra
data in order to achieve an emitting radius of the cooling component
consistent with the neutron star radius (otherwise the inferred radius would
be smaller; Rutledge et al. 2001a, 2001b).
The nature of this hard component is still uncertain. Models range from
Comptonization to Advection/Convection Dominate Accretion Flow (ADAF/CDAF) to
shock emission from the neutron star that resumed its radio pulsar activity in
quiescence. 
The latter model envisages a situation similar to that of the eclipsing
radio pulsar PSR B1259--63 or of the `black widow' pulsar PSR B1957+20:
a shock at the boundary between the relativistic MHD wind from the radio
pulsar and the matter outflowing from the companion star 
(Tavani \& Arons 1997; Tavani \& Brookshaw 1991;
Campana et al. 1998a). For the model to explain the observed luminosity in the
hard power law component of quiescent SXRTs, some few percent
of the pulsar spin-down luminosity must be converted into shock
emission\footnote{Ongoing deep searches in the radio band have not yet
revealed any steady or pulsed emission from quiescent SXRTs (Burgay et
al. 2003); however free-free absorption due to matter in the
binary system might be an important limiting factor in these searches (Stella
et al. 1994; Burgay et al. 2003).}. 
The shock emission model predicts synchrotron emission with power law photon
indexes in the 1.5--2 range and extending over a wide range of frequencies.
Indirect indications for the presence of this emission also from UV
observations of Cen X-4 with HST revealing a flat spectrum (i.e. $\Gamma\sim
2$) which matches well the extrapolated X--ray power law component (McClintock
\& Remillard 2000). Power law indexes outside the above range are indications
of strong (inverse-Compton) cooling. 

In a way similar to what is routinely done in the optical, a promising tool
to probe the X--ray emitting regions is through a rough eclipse mapping technique
(e.g. Horne 1985). Chandra observations of the eclipsing SXRT 4U 2129+47
were the first to exploit the potential of this technique by looking at the
extension of the emitting regions through eclipses (Nowak, Heinz \& Begelman
2002). During eclipses the soft component gets totally eclipsed whereas the
hard component is too faint to be revealed. This implies an upper limit on the
emission size of $\lsim 10\%$ the orbital separation (Nowak et al. 2002). 
The inclination of Aql X-1 has been estimated, from ellipsoidal
variations in the R and I band lightcurves, to be greater than 36 degrees
(Welsh, Robinson \& Young 2000).

In this paper we investigate in more detail the quiescent spectrum of one 
of the best studied SXRT sources: Aql X-1. We take advantage of four Chandra
exposures (Rutledge et al. 2002a) and one 76 ks long (unpublished) BeppoSAX
exposure. All these data were collected just after the November 2000 outburst. 
Based on the Chandra data Rutledge et al. (2002a) claimed that the soft
component decreased by $\sim 50\%$ over three months, then increased by $\sim
35\%$ in one month, and then remained constant ($<6\%$ change) over the last 
month. The variability of these observations was ascribed to an intrinsic
variability of the soft component hinting to accretion onto the neutron star
surface. Here we discuss in more detail these observations together with the
BeppoSAX long exposure, probing the shock emission model. 

In Section 2 we deal with the data. In Section 3 we describe the spectral
fitting and related results. Discussion and conclusions are reported in
Section 4.

\section{Data}

\subsection {Chandra}

Aql X-1 was observed by Chandra after the November 2000 outburst on four
occasions (see Table \ref{obs}). Observations were carried out with the
backside-illuminated ACIS-S detector (S3) at an off-axis position of $4'$ and
limited read-out area of 1/8 (achieving a time resolution of 0.44 s) in order
to limit problems connected to pile-up (see Rutledge et al. 2002a).
As expressly required, Aql X-1 fell on the same physical
pixels in order to avoid problems with the CCDs quantum efficiency (see
Rutledge et al. 2002a). For the analysis we use CIAO~2.2.1 with CALDB~2.15
(these are later versions than what used by Rutledge et al. 2002a).  
In all observations, we extracted the source counts from an elliptical 
$4.5''\times 3''$ region centered on source with a position angle matching the 
source. Background photons were extracted from an annular region with inner
and outer radii of $10''$ and $20''$, respectively.
Data were extracted, using {\em psextract}, into pulse-invariant (PI)
spectra. We grouped all the spectra to have (at least) 30 photons per channel.
We also corrected all the ancillary (arf) files with the recently released
{\em corrarf} tool to account for the continuous degradation in the ACIS
CCD's quantum efficiency.

\subsection{BeppoSAX}

We analysed data from the two imaging instruments on board the BeppoSAX
satellite: the Low Energy Concentrator Spectrometer (LECS; 0.1--10 keV,
Parmar et al. 1997) and the Medium Energy  Concentrator Spectrometer
(MECS; 1.6--10.5 keV, Boella et al. 1997). Non-imaging instruments provided
only upper limits. 
Only two of the three MECS units were operating at the time of the observations.
LECS data were collected only during satellite night-time resulting in shorter
exposure times. For a summary of the observations see Table \ref{obs}.
The observation took place on 2001 April 14, observing Aql X-1 for a
net exposure time of 76 ks with the MECS and 30 ks for the LECS.

Products were extracted using the FTOOLS package (v. 5.1). LECS and MECS
events were extracted from a circle of $4'$ radius. The background was
subtracted using spectra from blank sky files at the same detector coordinates
(after checking that the background of the observation was comparable).
We rebinned the LECS and MECS spectra in order to have 80 counts per spectral 
bin each.

\section{Overall spectral analysis}

\subsection{Spectral model}

Rutledge et al. (2002a) analysed the four Chandra spectra together. Spectra are 
different, e.g. a hard power-law tail has been detected only during two of the
four observations. To account for these differences Rutledge et al. (2002a)
considered a model made by an absorbed atmosphere model plus a power law and
let vary single parameters (fixed all the other) trying to account for the
variations. They found that differences in the spectra cannot be explained as
entirely due to either a changing power law flux and/or index or to a variable
column density.  On the contrary, differences can be acceptably (in terms of
$\chi^2$ statistics) explained as entirely due to a (non monotonic)
temperature variability of the thermal emission component. These variations
cannot be explained within the deep crustal heating model and the authors
suggested that these might originate from quiescent accretion onto the neutron
star surface.

As discussed in the introduction, we want to test here the hypothesis that
quiescent emission of SXRTs (and Aql X-1 in particular) are produced by a soft 
thermal component, likely arising from cooling of the neutron star, plus a
power law hard tail, arising from shock emission due to an active millisecond
pulsar. In theory, the soft component is steady on relatively short time
whereas the hard component can likely vary depending on the geometry and
density the outflowing matter. Hydrodynamical simulations (Brookshaw \& Tavani
1993) as well as radio observations of millisecond radio pulsars (MSPs) in
binary systems with a sizeable mass transfer show complex geometries and, more
importantly, variations from one orbital cycle to another. 
The example of the recently discovered MSP PSR J1740--5340 (D'Amico et
al. 2001; Ferrario et al. 2001) is enlighting. This is a 3.7 ms MSP orbiting
every 32.5 hr a main sequence companion. The source is located in the
globular cluster NGC 6397 at 2.5 kpc.
The radio pulsar gets partially and totally eclipsed over a wide range of orbital
phases. It emits X--rays as observed by Chandra (Grindlay et al. 2001) likely
arising from shock emission.
As testified by this source, the MSP is eclipsed for large part of the orbit
and variations from orbit to orbit are seen. 

This case motivates us to consider a spectral model for fitting the quiescent
X--ray spectra of Aql X-1 made by a soft thermal component from the entire
neutron star, a variable power law component (the strength of which depends on
the interaction with the surrounding matter) {\it and} a variable column
density due to variations intrinsic to the source (over a fixed interstellar
column density). 

\bigskip

\subsection{Spectral analysis}

\begin{figure*}
\begin{center}
\psfig{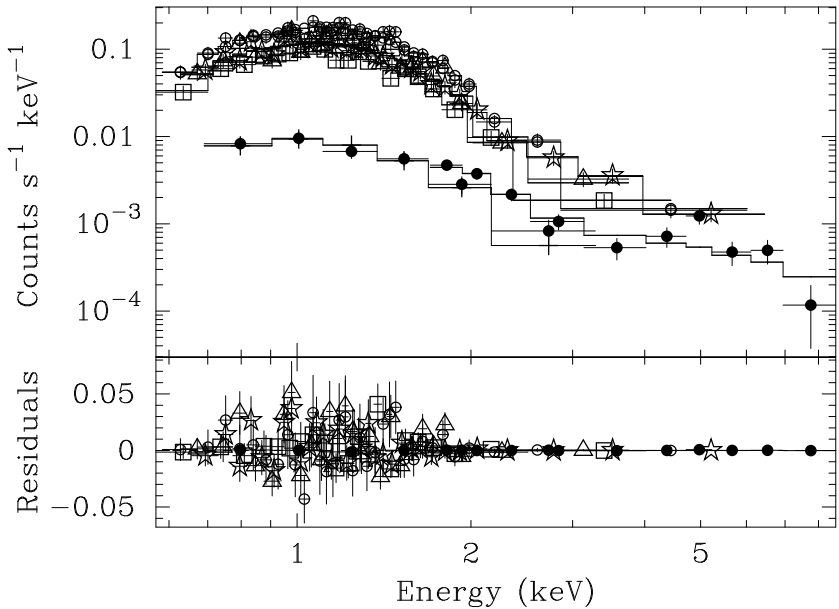}
\caption{Aql X-1 spectra of the five observations described in the text. The
Chandra spectra are in the upper part of the figure. LECS and MECS spectra are 
indicated with filled circles. The best fit model is overlaid on the data.}
\label{fig1}
\end{center}
\end{figure*}

Spectral analysis was carried out with the XSPEC software (v. 11.2.0). 
The spectral model we adopted consist of a (fixed) cooling spectrum (we use 
here the hydrogen atmosphere model by G\"ansicke, Braje \& Romani 2002, {\tt
hyd\_spectra.mod} in XSPEC) plus a variable power law. A variable column
density was also adopted ({\tt tbabs} in XSPEC, Wilms, Allen \& McCray 2000).
We fitted all the five spectra (4 Chandra and 1 BeppoSAX, LECS plus MECS)
together. For the BeppoSAX data we used the public response matrices available
in January 2000. During the fit a variable normalisation factor between LECS
and MECS was included to account for the mismatch in the absolute flux calibration
of the BeppoSAX instruments (Fiore, Guainazzi \& Grandi 1999).
A variable normalisation factor was also included to account for mismatch
between BeppoSAX and Chandra\footnote{We find a ratio of the BeppoSAX over the 
Chandra normalization of $1.9^{+1.4}_{-0.6}$, consistent with previous
determinations (e.g. Piro et al. 2001).}. The spectrum provides a
statistically acceptable description of the entire data set with a reduced
$\chi^2$ of 1.00 (null hypothesis probability $49.0\%$, see Table \ref{spe}
and Fig. \ref{fig1}).  
This indicates that, at least at a first sight, all the data concerning the
quiescence following the November 2000 outburst of Aql X-1 are consistent with 
a model made by a cooling neutron star and a variable power law and absorption 
components. 

\begin{figure*}
\begin{center}
\psfig{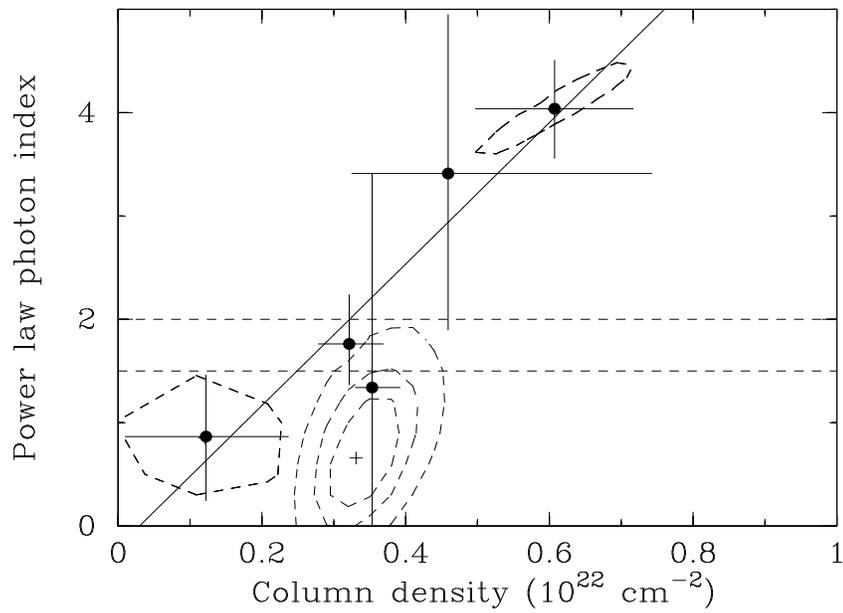}
\caption{Power law photon index vs. column density correlation of the five Aql
X-1 observations. Overplotted is the best linear fit. Dashed lines indicate
the range over which the synchrotron emission model likely applies. On the
hardest and softest observations $1\,\sigma$ countours have been superposed. 
The 1, 2, $3\,\sigma$ countours obtained fitting the entire set of data with a 
single power law and column density is also reported.}
\label{fig2}
\end{center}
\end{figure*}

A detailed theory of the shock emission mechanism has been developed by Tavani 
\& Arons (1997) tailored on the young radio pulsar PSR B1259--63 orbiting a Be 
star. A more detailed discussion on SXRTs is within Campana et
al. (1998a). The expected spectrum is a power law spectrum with photon index
in the 1.5--2 range, with a positive correlation between the quantity of matter
at the shock region (possibly traced by the column density) and the power law
index. This behaviour has been observed in PSR B1259--63 which showed a photon
index of $\sim 2$ at periastron and hardened towards apastron.
However, the column density to PSR B1259-63 is larger than the one
to Aql X-1 and we expect that variations in the column density are therefore
highly suppressed. 
This correlation might provide us with a further check on the shock emission
mechanism. 
In Fig. \ref{fig2} we show the column density and power law indexes for the
five observations. A correlation is indeed present suggesting that this
mechanism might be at work. We note however that the column density and the
power law index are correlated parameters and for this reason we
derived the errors ($68\%$ c.l. in Fig. \ref{fig2}), for the two parameters
together ($\Delta \chi^2=2.30$). Correlation is tight. A fit with a constant
provides a  $\chi^2_{\rm red}=5.4$ with a null hypothesis probability of
$2\times 10^{-4}$ whereas a linear fit gives $\chi^2_{\rm red}=0.2$ (with an
F-test probability of $99.8\%$). 
We also carried out a weighted linear correlation test finding a
correlation probability of $r_{\rm w}=0.9$ ($95\%$ probability).
A further correlation can be tested between the power law flux
and the column density. A weighted Pearson correlation test gives $r=0.8$ 
($90\%$ probability).

A similar correlation has also been found in the quiescent emission of
the transient black hole candidate V404 Cyg (Kong et al. 2002). They however
suspected that the correlation is not intrinsic to the source, but it is 
an artifact of the fitting process. In their case the slope of the correlation
is nearly (to within $5\%$) the same as the slope of the major axis of the
parameter confidence contours. Moreover, they did not found correlation
between flux and column density. This is contrary to our findings. 
In Fig. 2 we also included the confidence countors showing that this
alignement effect is not present.

We consider then a cut-off power law model in addition to the neutron
star atmosphere one. The overall fit is as good as the others ($\chi_{\rm
red}^2=1.05$, null hypothesis probability of $35\%$). The cut-off energy is
only loosely constrained and only an upper limit for each observation can be
put. Drawing contour plots in the cut-off energy -- column density plane we
find an anticorrelation between the two quantities: the lower the energy of
the cut-off the larger the column density. From a Pearson correlation test we
obtain $r=-0.9$, corresponding to a significance of $\sim 96\%$. 
This might indicate an (almost) constant luminosity that is shared among
different quantities of matter.
 
To further test this idea we fit the data with a different model with a smaller
emission at low energies. We consider a model consisting of a Comptonized 
component ({\tt COMPTT}, Titarchuk 1994) plus an atmosphere component.
We found a large degree of freedom with this model which forced us to freeze
the input soft photon temperature to the temperature of the neutron star
atmosphere (that is the same for all the observations) and take the same
amount of optical depth for all the observations (that turns out to be 7.7)
and leave the plasma temperature free to vary. The fit is as good as the one
with the power law ($\chi_{\rm red}^2=1.00$). 
We derive an anti-correlation between the column density
and the plasma temperature. 
A weighted Pearson correlation test gives $r=-0.9$ ($\sim 93\%$ probability).

\section{Discussion}

Rutledge et al. (2002a) analysing Chandra data of the Aql X-1 quiescent phase 
after the November 2000 outburst found a variable flux and X--ray spectrum.
They interpreted these variations in terms of variations of the neutron star 
effective temperature, which changed from $130^{+3}_{-5}$ eV (C1), down to
$113^{+3}_{-4}$ eV (C2), and finally increased to $118^{+9}_{-4}$ eV (C3, C4).
Interestingly,  during observation C4 they also found short-term variability
(at $32\%$ rms) and a possible absorption feature near 0.5 keV (even if this 
feature can also be explained as due to a time-variable response in the ACIS
detector; Rutledge et al. 2002a). 

Short term variability is a powerful tool for the study of the emission
mechanism(s) responsible for the SXRTs quiescent emission. 
A factor of 3 variability over timescales of days (Campana et al. 1997) 
and $40\%$ over 4.5 yr (Rutledge et al. 2001b) has been reported in Cen X-4.
Several other neutron star systems have also been found to be
variable in quiescence by factors of 3--5 (e.g. Rutledge et al. 2000) but data
have been collected over several years and with different instruments.
Chandra data on Aql X-1 are the first that show a clear luminosity variation
and, more importantly, an increase during quiescence. No known mechanism
associated with crustal heating can account for this variability (Rutledge et
al. 2002a). 

Here we approach the same Chandra data plus an unpublished long BeppoSAX
observation of Aql X-1 in quiescence carried out in the same period to probe
a different spectral model. Deep crustal heating (Brown et al. 1998; Rutledge
et al. 1999; Colpi et al. 2001) has been proposed as a physically sound
mechanism powering the soft component of the quiescent spectra of SXRTs.
Several mechanisms have been proposed to explain the hard tail component often 
observed in quiescent SXRTs. One of them, physically motivated by the recent
observations of the MSP PSR J1740--5340 (D'Amico et al. 2001; Grindlay et
al. 2001), relies on the shock emission between the relativistic MSP wind and
matter outflowing from the companion (Tavani \& Arons 1997; Campana et
al. 1998a). These two components are not exclusive.
With this physical scenario, we fit the spectra fixing the soft component for
all the observations and leave free to vary the hard component {\it and} the
column density (that changes by a factor of $\sim 5$).
This model is consistent with the
entire dataset ($\chi^2_{\rm red}=1.00$, for 109 d.o.f. and with a null
hypothesis probability of $49.0\%$). Fitting the same dataset with the
best fit model by Rutledge et al. (2002a) with the addition of the BeppoSAX
data, we obtain a slightly worse fit ($\chi^2_{\rm red}=1.17$, for 113
d.o.f. and with a null hypothesis probability of $11.1\%$). 
An equally good fit is provided by a neutron star atmosphere plus a
Comptonization component ({\tt COMPTT}). This model provides a high degree of
freedom with a $\chi^2_{\rm red}=0.99$ fit (for 104 d.o.f. and with a null
hypothesis probability of $50.5\%$) obtained fixing the soft Wien 
temperature to atmosphere temperature and the plasma temperature kept the same
in all the observations.
We conclude that the scenario proposed is (at least) equally
well consistent with the data, meaning that a shock emission scenario can
account for the spectral variability observed in Aql X-1.
We also note that Rutledge et al. (2002) found $32\%$ (rms) variability in 
observation C4.  In their case the power law component contributed only $12\%$
of the flux. From our fit the power law component contributes to $38\%$ of the
total flux, so it can in principle account for all of the short-term
variability.  
 
Despite the low number of points, spectral parameters derived for the power
law index show some correlation with the column density (interpreted as a
measure of the variable mass around the system, over a fixed interstellar
amount) as well as with the power law flux. This correlation might be expected in 
the shock emission scenario (Tavani \& Arons 1997). 
What is now expected is the large value of the power law index in the last
observations. This might then provide an indication of a different regime in
the system, possibly underlying a larger inverse Compton cooling. The hard
part of the spectrum is in fact consistent also with a thermal bremsstrahlung
spectrum. 

Further observations can shed light on this new interesting field, namely
variability in the quiescent phase of SXRTs, which up to now has been often
unconsidered.

\begin{acknowledgements}
We thank an anonymous referee and G. Ghisellini for useful comments.
\end{acknowledgements}

\begin{table*}[!htb]
\begin{center}
\caption{\label{obs} Log of Aql X-1 observations.}
\begin{tabular}{ccccc}
\hline\hline
Satellite & Seq. Num.& Start Time & Exposure & Orbital phase   \\
          &          &            & (s)      & $\phi_{\rm orb}$\\
\hline
Chandra   &  400075  & 2000-11-28 &\ 6628    &0.02--0.15 (\ppm0.02)\\
Chandra   &  400076  & 2001-02-19 &\ 7787    &0.20--0.36 (\ppm0.02)\\
Chandra   &  400077  & 2001-03-23 &\ 7390    &0.19--0.34 (\ppm0.02)\\
Chandra   &  400078  & 2001-04-20 &\ 9245    &0.22--0.39 (\ppm0.02)\\
\hline
BSAX LECS & 212380011& 2001-04-14 & 30390    &0.38--0.50 (\ppm0.02)\\
BSAX MECS & 212380011& 2001-04-14 & 76301    &0.38--0.50 (\ppm0.02)\\
\hline\hline
\end{tabular}
\end{center}

\tablecomments{$^a$ Orbital phase relative to minimum light (inferior
conjunction of the secondary), ephemeris from Garcia et al. (1999). For
Chandra data these are taken from Rutledge et al. (2002a).}
\end{table*}

\begin{table*}[!htb]
\begin{center}
\caption{\label{spe} Spectral fit of Chandra and BeppoSAX Aql X-1
observations.} 
\begin{tabular}{ccc}
\hline\hline
Parameter     & Value              & Component flux   \\
              & ($90\%$ c.l.)      & ($10^{-12}$ cgs) \\
\hline
Temp. (eV)    & $157_{-32}^{+31}$  & \\
Radius (km)$^*$&$11.1_{-4.3}^{+8.2}$& 1.5\\                  
\hline
\hline
$N_H$ (S)     & $1.2_{-1.1}^{+1.3}$& \\
Power law (S) & $0.9_{-0.7}^{+0.6}$& 0.4 ($22\%$) \\         
\hline
$N_H$ (C1)    & $6.1_{-1.2}^{+1.2}$& \\
Power law (C1)& $4.0_{-0.5}^{+0.5}$& 6.2 ($81\%$) \\         
\hline
$N_H$ (C2)    & $3.5_{-0.5}^{+0.6}$& \\
Power law (C2)& $1.3_{-5.3}^{+2.3}$& 0.3 ($16\%$) \\         
\hline
$N_H$ (C3)    & $4.6_{-1.4}^{+3.1}$& \\
Power law (C3)& $3.4_{-1.6}^{+1.6}$& 1.5 ($51\%$) \\         
\hline
$N_H$ (C4)    & $3.2_{-0.4}^{+0.5}$& \\
Power law (C4)& $1.8_{-0.6}^{+0.5}$& 0.9 ($38\%$) \\         
\hline\hline
\end{tabular}
\end{center}

\noindent S indicates the BeppoSAX observations, C1 to C4 the four Chandra
observations. Column density values are in units of
$10^{21}\cmdue$. Confidence levels have been computed for one parameter of
interest at $90\%$ (i.e. $\Delta \chi^2=2.71$), this is at variance 
with discussed in the text and is motivated to allow a comparison between all 
the parameters.
Fluxes are unabsorbed and in the 0.5--10 keV energy band. Number in
parenthesis indicate the percentage of the total flux in the power law
component.

\noindent Observation C1 where the power law is steep can be equally
well fit with a bremsstrahlung model with temperatures $k\,T=2.0^{+2.4}_{-0.8}$
keV and $N_H=4.0^{+3.8}_{-3.2}\times 10^{21}\cmdue$.

\noindent $^*$ Temperature and radius are at the neutron star.
Radius at a distance of 4 kpc.

\end{table*}


\newpage

\end{document}